\documentclass[aps,pra,onecolumn,showpacs,twosides,superscriptaddress,floatfix,nofootinbib]{revtex4}
\usepackage{amsmath}
\usepackage{amsfonts}
\usepackage{amssymb}
\usepackage{graphicx}
\usepackage[ps2pdf,bookmarks=true]{hyperref}
\usepackage{mathrsfs}

\begin{document}

\title{Quasi-Hamiltonian Method for Computation of Decoherence Rates}
\author{Robert Joynt}
\email{rjjoynt@wisc.edu}
\author{Dong Zhou}
\email{zhou.dong@gmail.com}
\affiliation{Department of Physics, University of Wisconsin-Madison, 1150 Univ. Ave., Madison, WI 53706, USA}

\author{Qiang-Hua Wang}
\email{qhwang@nju.edu.cn}
\affiliation{National Laboratory of Solid State Microstructures and Department of Physics, Nanjing University, Nanjing 210093, China}

\date{\today}

\begin{abstract}
We present a general formalism for the dissipative dynamics of an arbitrary
quantum system in the presence of a classical stochastic process. It is
applicable to a wide range of physical situations, and in particular it can be
used for qubit arrays in the presence of classical two-level systems (TLS).
In this formalism, all decoherence rates appear as eigenvalues of an
evolution matrix. Thus the method is linear, and the close analogy to
Hamiltonian systems opens up a toolbox of well-developed methods such as
perturbation theory and mean-field theory. We apply the method to the
problem of a single qubit in the presence of TLS that give rise to pure
dephasing 1/f noise and solve this problem exactly. The exact solution gives
an experimentally observable improvement over the popular Gaussian
approximation. 
\end{abstract}


\maketitle
\section{Introduction}

With researchers motivated by the prospect of quantum computing, qubit
dephasing has been a topic of intense research over the past decade. Various
models of this venerable \cite{klauder} phenomenon have been investigated.
 The most popular have been the spin-boson or spin bath models \cite{leggett,
 weiss, schoen,stamp}.  Another important version has
been that of an electron spin coupling to nuclei \cite{glazman,sarma}.
 In recent studies of superconducting qubits, however, it has been found that
1/f-type noise is the chief source of dephasing \cite{sendelbach,
yoshihara,kakuyanagi,bialczak,wellstood}.  The
sources of this noise are two-level systems (TLS) \cite{halperin,
phillips,weissman} with a wide spectrum of switching rates.  It
is likely to be important in virtually any solid-state system that serves as a
host for qubits, as TLS are ubiquitous in bulk materials.  This noise is
usually modeled as classical noise.  

Our aim in this paper will be to present a formalism that solves for the
dissipative dynamics of an arbitrary quantum system in the presence of a
classical stochastic process.  This is a very general model of classical
noise.  The formalism depends on a combination of the "Liouvillian" approach
to the evolution of the density matrix \cite{blum} with methods from the
classical theory of stochastic processes \cite{van kampen}. In particular, the
formalism applies to an ensemble of TLS with any distribution of switching
rates and couplings to the quantum system.  It has the great advantage of
reducing to a linear system of equations, and in fact there is a close analogy
to the usual Hamiltonian formulation of quantum mechanics.  It is exact,
making no approximation as to the strength of the coupling relative to the
inverse of the time scales of the noise. 

This method has been derived for a specific example in previous work
\cite{cheng}.  As an illustrative case, we use the new method to solve the
problem of a single qubit in the presence of TLS that give rise to pure
dephasing 1/f noise.  Other solutions of this problem have been found by
previous authors \cite{galperin, paladino}, there have been numerical
studies \cite{faoro}, and the subject has recently been comprehensively
reviewed \cite{bergli}, so this problem is a good testbed for our method.  It
also allows us to exhibit the Hamiltonian analogy, which in this case is to a
spin 1/2 system.  The illustrative case points the way to other interesting
models that are not exactly solvable, but to which the method also applies.

This paper is concerned with mathematical methods.  Application to specific
physical systems will be given in future work.  The particular case of
superconducting qubits has recently been treated \cite{zhou}.  The main new
results of a general nature are found in Eqs. \ref{eq:gamma}, \ref{eq:Hq} and
the physical interpretation following Eq. \ref{eq:interpretation}.  New
results for strong-coupling (1/f and similar) noise are found in Eqs.
\ref{eq:strongt} and \ref{eq:manystrong}.  The most convenient starting point
for future calculations of the effects of 1/f and other broad-spectrum noise
is found in Eq. \ref{eq:bsnstart}. 

\section{General Method}

We consider the general problem of a quantum system in the presence of
classical noise.  The quantum system is an $N_{q}$-state system, so its
Hilbert space is $N_{q}$-dimensional.  The classical system has $N_{c}$
states labeled by the index $a$.  The initial state of the composite system
is given by the Hermitian $N_{q}\times N_{q}$ density matrix $\rho\left(
t=0\right)  $ and the classical probability distribution $P_{a}\left(
t=0\right)  $.  $0\leq P_{a}\leq1$.  $\rho$ and $P$ satisfy
\begin{equation}
\text{Tr }\rho\left(  t\right)  =1\text{ and }\sum_{a=1}^{N_{c}}P_{a}\left(
t\right)  =1
\end{equation}
at all times $t$.  The classical environment passes through a sequence $S$ of
discrete states during the course of the time evolution.  The probability
distribution of these states evolves according to the master equation
\begin{equation}
\frac{dP_{a}\left(  t\right)  }{dt}=\sum\limits_{b=1}^{N_{c}}V_{ab}
P_{b}\left(  t\right)  . \label{eq:master}
\end{equation}
$V$ is a real matrix of transition probabilities.  It satisfies $\sum
_{a}V_{ab}=0$.  The Hamiltonian for the quantum system is $H$: it is a
function of the sequence of states of the classical environment, and is
therefore time-dependent.  For a fixed sequence $S$ the density matrix
evolves according to the Von Neumann equation
\begin{equation}
\frac{d\rho_{S}}{dt}=-i\left[ H\left(  S\right) , \rho_{S} \right]
\label{eq:vn}
\end{equation}
in units with $\hbar=1$.  However, we are interested in the density matrix
averaged over all sequences.  We shall denote averages over $S$ by an
overbar, so the actual density matrix is $\rho=\overline{\rho_{S}}$.  We
shall treat both the quantum system and the classical environment as
finite-dimensional. 

Since $a$ is a classical random variable, this is a classical noise model.
 The model applies when the noise sources are more strongly coupled to an
external bath than to the qubit, so that back action of the qubit on the noise
sources is negligible.  The Hamiltonian $H$ is a function of $a$, the state
of the classical system, but $V_{ab}$ is independent of $\rho$.  This implies
that quantum information that leaves the qubit leaves forever.  The
conditions under which such a model is appropriate have been considered in
more detail by Galperin \textit{et al}. \cite{galperin}.  

\subsection{Transfer Matrix for a Fixed Noise Sequence}

We wish to compute the qubit density matrix $\rho_{S}\left(  t\right)$,
given $\rho_{S}\left(  0\right)$, for a fixed sequence $S$.  Our first step is to rewrite this in
terms of the evolution of a generalized Bloch vector $n_{i}\left(  t\right)$:
\begin{equation}
\rho_{S}\left(  t\right)  =\frac{1}{N_{q}}\left[I+\sum_{i=1}^{N_{q}^{2}-1}
n_{i}\left(  S,t\right)  \lambda_{i}\right], \label{eq:expansion}
\end{equation}
where $n_{i}$ is a set of $N_{q}^{2}-1$ real numbers, $I$ is the $N_{q}\times
N_{q}$ unit matrix and $\lambda_{i}$ are the generators of $SU\left(
N\right)$.  The $\lambda_{i}$ are time-independent $N_{q}\times N_{q}$
matrices and they are chosen to satisfy
\begin{equation}
\text{Tr }\lambda_{i}=0,\text{ }\lambda_{i}^{\dagger}=\lambda_{i},\text{ and
}\frac{1}{2}\text{Tr }\lambda_{i}\lambda_{j}=\delta_{ij}.\label{eq:tr}
\end{equation}
The $\lambda_{i}$ form an orthonormal basis for the quantum state space of
density matrices under the inner product $\left(  \rho_{1},\rho_{2}\right)
=\frac{1}{2}$Tr ($\rho_{1}\rho_{2}).$ The fact that the $\lambda_{i}$ are
traceless, together with Eq. \ref{eq:expansion} , immediately implies the
conservation of probability: $\text{Tr}\rho=1$.

Consider a short time interval $\Delta t$ in which $H$ is constant and the
environment is in a fixed state $a$.  The formal solution to Eq. \ref{eq:vn}
is
\begin{equation}
\rho_{S}\left(  a,\Delta t\right)  =U\left(  a,\Delta t\right)  \rho\left(
0\right)  U^{\dag}\left(  a,\Delta t\right)  .\label{eq:ul}
\end{equation}
with $U\left(  a,t\right)  =\exp\left[  -itH\left(  a\right)  \right]  $. In
terms of the $\lambda_{i}$, Eq. \ref{eq:ul} is
\[
\frac{1}{2}I_{\alpha\beta}+\frac{1}{2}n_{i}\left(  a,\Delta t\right)
~\lambda_{i},_{\alpha\beta}=U_{\alpha\gamma}\left(  a,\Delta t\right)
U_{\delta\beta}^{\dag}\left(  a,\Delta t\right)  \left[  \frac{1}{2}%
I_{\gamma\delta}+\frac{1}{2}n_{i}\left(  0\right)  \lambda_{i},_{\gamma\delta
}\right]  .
\]
where we have temporarily included Greek subscripts for clarity.  These
indices denote components in the Hilbert space of the quantum system.  They
take on the values $\alpha=1,2,\ldots,N_{q}$. The Roman subscripts take on the
values $i=1,2,...,N_{q}^{2}-1$.  Both are subject to a summation convention.
\ The identity matrix term cancels out ($I_{\alpha\beta}$ has no dynamics) and
we have
\[
n_{i}\left(  a,\Delta t\right)  \lambda_{i,\alpha\beta}=U_{\alpha\gamma
}\left(  a,\Delta t\right)  U_{\delta\beta}^{\dag}\left(  a,\Delta t\right)
~n_{i}\left(  0\right)  \lambda_{i,\gamma\delta}.
\]
We may extract the components of $n$ by multiplying this equation by the
$i$-th generator and taking the trace over the Greek indices.  Using the
trace identity from Eq. \ref{eq:tr} we find
\[
n_{i}\left(  a,\Delta t\right)  =\frac{1}{2}U_{\alpha\gamma}\left(  a,\Delta
t\right)  U_{\delta\beta}^{\dag}\left(  a,\Delta t\right)  ~n_{j}\left(
0\right)  \lambda_{j,\gamma\delta}~\lambda_{i,\beta\alpha}.
\]
This is conveniently written as
\[
n_{i}\left(  a,\Delta t\right)  =T_{ij}\left(  a,\Delta t\right)  n_{j}\left(
0\right)  ,
\]
where
\begin{equation}
T_{ij}\left(  a,\Delta t\right)  =\frac{1}{2}\text{Tr~}\left[  \lambda
_{i}U\left(  a,\Delta t\right)  \lambda_{j}U^{\dag}\left(  a,\Delta t\right)
\right]  .\label{eq:tij}
\end{equation}
$T_{ij}\left(  a,\Delta t\right)  $ is the quantum dynamical map (sometimes
referred to as the Liouvillian) for the interval $\Delta t$ in an environment
in state $a$ (Some properties of $T$ are given in App. \ref{sec:app_A}). 
From now on the Greek indices will be suppressed; operations in
the quantum Hilbert space of operators are indicated by matrix multiplication
and the trace. 

Thus for the whole sequence $S$,
\begin{align}
\mathbf{n}_S(t) = T(a_N)\cdots T(a_2)~T(a_1)~\mathbf{n}(0),
\end{align}
where $t=N\Delta t$ and $a_k$ labels the state of the classical environment in the time interval $k\Delta t$.

\subsection{Averaging over All Noise Sequences}
We are interested in the generalized Bloch vector averaged over all possible sequences,
i.e. $\mathbf{n}(t) = \overline{\mathbf{n}_S(t)}=\overline{T}(t)\mathbf{n}(0)$.

\begin{figure} [tbp] 
\begin{center}
\includegraphics*[width=0.8\linewidth] {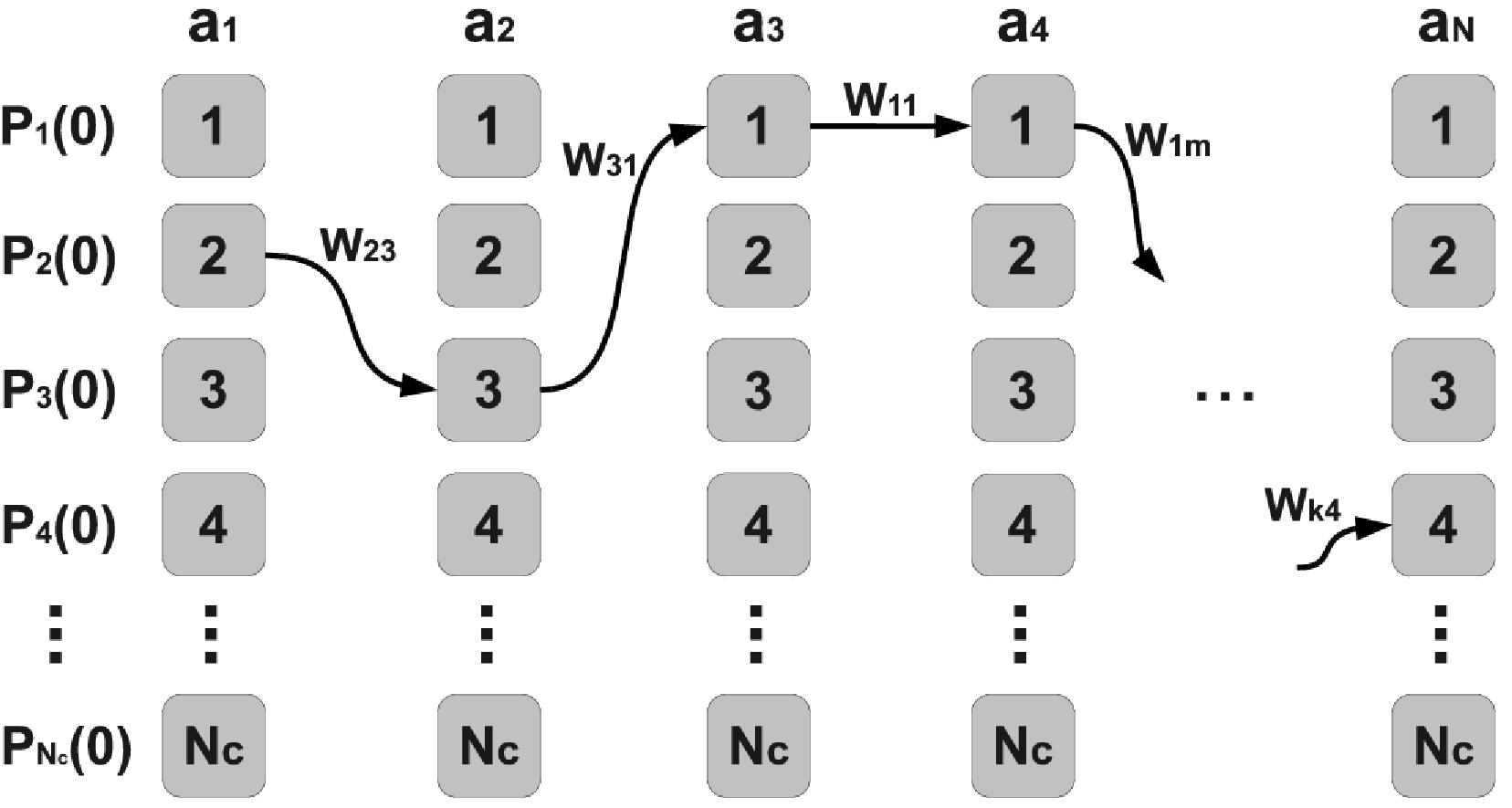}
\caption{A fixed sequence $S=\{2,3,1,1,m,\ldots,k,4\}$ of the classical noise. $P_{i}(0)$ are the initial probability distributions of the $N_c$ states. $W_{a_i,a_{i+1}}$ are the infinitesimal transition probability between state $a_i$ and $a_{i+1}$ at the instant $i\Delta t$.}
\label{fig:path}
\end{center}
\end{figure}

To compute $\overline{T}(t)$, we note each noise sequence $S=\{a_1,a_2,\ldots,a_N\}$ is associated with probability $W_{a_0,a_1}W_{a_1,a_2}\cdots W_{a_{N-1},a_N}$, 
where $W_{a_k,a_{k+1}}$ is the transition probability of the classical system going from state $a_k$ to $a_{k+1}$ at the end of the $k$'th time interval, see Fig. \ref{fig:path}  and the infinitesimal expansion of $W(\Delta t)$ is given by 
\begin{align}
W(\Delta t) = I_c + V\Delta t
\end{align}
where $I_c$ is the $N_c\times N_c$ unit matrix.
$W_{a_0,a_1}$ is put in by hand for later on convenience. As $N\rightarrow\infty$ it will not introduce any errors.

Let us defined a $[N_{c}(N_{q}^{2}-1)]\times[N_{c}(N_{q}^{2}-1)]$ tensor $\mathbf{\Gamma}$  whose element  is
\begin{equation}
\boldsymbol{\Gamma}(a_{r},a_{r-1})=W_{a_{r},a_{r-1}}
\otimes{T}(a_r). \label{eq:gamma}
\end{equation}

The averaged Bloch vector is thus given by
\begin{align}
\mathbf{n}(t) =  \sum_{a_1,a_2,\ldots,a_N}P_{a_1}(0)\mathbf{\Gamma}(a_N,a_{N-1})\cdots\mathbf{\Gamma}(a_1,a_0)\mathbf{n}(0)
\end{align}

Equivalently, we can utilize the tensor nature of $\mathbf{\Gamma}$ and write
\begin{align}
\mathbf{n}(t) = \left<x_f\right|\mathbf{\Gamma}^N\left|i_f\right>\mathbf{n}(0)
\end{align}
where $\left<x_f\right|=[1,1,\ldots,1]$ and $\left|i_f\right>=[p_1(0),p_2(0),\ldots,p_{N_c}(0)]$ act on the clssical environment. 
This contraction amounts to averaging over all the $N_c\times N_c$ blocks of $\mathbf{\Gamma}^N$, 
each of which corresponds to the family of time evolutions caused by noise sequences that start from $a_1$ and end with $a_N$.
In this formalism, it is possible to put the classical environment into an arbitrarily initial state $\left|i_f\right>$. 
 However, in almost all cases of physical interest, both the initial and final states of
the environment will be the stationary distribution $\left\vert p_{s}
\right\rangle $: the right eigenvector of $V$ corresponding to the eigenvalue
zero \cite{van kampen}.

This repeated matrix multiplication structure is the key to the formalism. Expand the matrix $\Gamma$ as
\begin{equation}
\Gamma(\Delta t) \approx I-iH_{q}\Delta t
\end{equation}
so that in the limit $\Delta t\rightarrow0$ with $t=N\Delta t$ held fixed we
have
\begin{equation}
\Gamma^{N}(t)=\left(I -iH_{q}\Delta t\right)^N =\exp\left(-iH_{q}t\right)
,\label{eq:quasi}
\end{equation}
and 
\begin{align}
\mathbf{n}(t) = \overline{T}(t)\mathbf{n}(0)= \left<x_f\right|e^{-iH_q t}\left|i_f\right>\mathbf{n}(0).
\end{align}
$H_{q}$ is the \textit{time-independent} "quasi-Hamiltonian" given by
\begin{equation}
H_{q}=i\lim_{\Delta t\rightarrow0}\frac{\Gamma(\Delta t)-I}{\Delta t}=i\frac{d}{dt}\Gamma(t=0).\label{eq:Hq}
\end{equation}
It completely characterizes the evolution of the open quantum system.
 $H_{q}$ is pure imaginary and non-Hermitian.  The idea of
a non-Hermitian Hamiltonian to characterize dissipation or absorption in
quantum systems is not new, going back at least to the optical model of the
nucleus \cite{mayer}.  However, the implementation has historically been
phenomenological; the results here are exact.

We write the eigendecomposition of $H_{q}$ as
\[
H_{q}=\sum\limits_{\psi}~\left\vert \psi\right\rangle ~\omega_{\psi
}~\left\langle \psi\right\vert ,
\]
with $\left\vert \psi\right\rangle $ and $\left\langle \psi\right\vert $ being
the right and left eigenvectors of $H_{q}.$  Note that since $H_{q}$ is not
Hermitian, $\left\vert \psi\right\rangle $ and $\left\langle \psi\right\vert $
are not dual to each other as in ordinary quantum mechanics. 

Note $\left\vert \psi\right\rangle $ is a state of the combined environment-qubit
system and the total evolution is given by
\begin{equation}
e^{-iH_{q}t}=\sum\limits_{\psi}\left\vert \psi\right\rangle e^{-i\omega_{\psi
}t}\left\langle \psi\right\vert , \label{eq:interpretation}
\end{equation}
so that $\operatorname{Re}\left(  \omega_{\psi}\right)  $ gives the
oscillation frequencies and $-\operatorname{Im}\left(  \omega_{\psi}\right)  $
gives the decay rates of the combined system.  $\psi=1,2,...,\left[
N_{c}\left(  N_{q}^{2}-1\right)  \right]$.  Included in this list of $-\operatorname{Im}(\omega_\psi)$'s are the rates for the environment. 

This formalism provides a means of calculating the dissipative evolution of
any quantum system evolving in the presence of a classical stochastic process.
 Furthermore it is completely linear, which means that all of the 
powerful techniques of linear algebra can be brought to bear, including
well-controlled perturbation theory.  In favorable cases such as the one to
be considered next, the quasi-Hamiltonian is similar to Hamiltonians familiar
from other problems in classical or quantum mechanics.  The arsenal of
methods developed for these situations can then be brought to bear.

We note that the derivation of the quasi-Hamiltonian is similar to the
time-slice derivation of the path integral approach to open quantum systems, which
leads to the Feynman-Vernon formulas for the influence functional
\cite{feynman}.  The difference in starting points is that the environment
here is taken as classical.  The difference in end results is quite
startling, since a non-Hermitian Hamiltonian does not seem to emerge naturally
from the Feynman-Vernon approach in any obvious limit.

We note finally that the present formalism is also ideal for the investigation
of quantum control schemes such as pulsing a qubit.  We only need to sandwich
the pulsing operators between the evolution operators.  Let the pulsing
operator be a unitary operator $R$ that acts at the time $t_{p}$ with $0<t_{p}<t$. 
Its action on the generalized Bloch vector is given by the
$\left(  N_{q}^{2}-1\right)  \times\left(  N_{q}^{2}-1\right)  $ matrix
\begin{equation}
\mathbf{R}_{ij}=\frac{1}{2}\text{Tr }\lambda_{i}R\lambda_{j}R^{\dag}
\end{equation}
and writing $\mathbf{U}_{p}=I_{c}\otimes\mathbf{R},$ we have
\[
\mathbf{n}\left(  t\right)  =\left<x_f\right|e^{-iH_{q}\left(
t-t_{p}\right)}\mathbf{U}_{p}e^{-iH_{q}t_{p}}\left|i_f\right>~\mathbf{n}\left(  0\right) .
\]
The generalization to more complicated pulsing schemes is immediate.

\section{Qubit Dephasing by Two Level Systems}

\subsection{Quasi-Hamiltonian}

We now proceed to solve exactly the problem of the evolution of the density
matrix of a qubit in the presence of an environment of $M$ independently
fluctuating TLS that dephase the qubit.  Even this simple case of $N_{q}=2$
and $N_{c}=2^{M}$ is of great experimental interest.  From the exact formulas
we will derive qualitative information by extracting and analyzing asymptotic
expressions in various limits. 

For $M$ statistically independent TLS we can describe the state of the
environment by variables $s_{n}\left(  t\right)  =\pm1$ that switch at random
intervals at an average rate $\gamma_{n}$.  $n=1,2,...,M$.  The most general
expression for the flipping probability matrix is $W=W_{1}\otimes\cdot
\cdot\cdot\otimes W_{M}$ with
\begin{equation}
W_{n}=
\begin{pmatrix}
1-p_{n}-\delta_{n} & p_{n}-\delta_{n}\\
p_{n}+\delta_{n} & 1-p_{n}+\delta_{n}
\end{pmatrix}
, \label{eq:wmatrix}
\end{equation}
or, in index notation $\left(  W_{n}\right)  _{+1,+1}=1-p_{n}-\delta_{n}$,
etc.  This states that the probability of starting and finishing the interval
in the $+1$ state is $1-p_{n}-\delta_{n},$ the probability of starting in the
$+1$ state and ending in the $-1$ state is $p+\delta,$ etc.  We can then
write $W_{n}=1-p_{n}+p_{n}\tau_{nx}-\delta\tau_{nz}-i\delta\tau_{ny}$, where
$\tau_{ni}$ are the Pauli matrices that act in the state space of fluctuator
$n$. The switching rate is $\gamma_{n}=p_{n}/\Delta t$, while $\delta_{n}$
controls the average occupation of the states.  We shall focus on the case
$\delta_{n}=0$, the unbiased fluctuators, when we have
\begin{equation}
W_{n}=\left(  1-p_{n}\right)  I+p_{n}\tau_{nx}.\label{eq:w1}
\end{equation}
and the stationary state is $\left|p_s\right>=[1, 1, \ldots, 1]/2^M$, which is the unbiased fluctuator.

Here $N_{q}=2$ and the generators of $SU(2)$ are the Pauli matrices
$\sigma_{x},\sigma_{y}$, and $\sigma_{z}$.  The Hamiltonian of the quantum
system is
\begin{equation}
H\left(  t\right)  =-\frac{1}{2}B_{0}\sigma_{z}-\frac{1}{2}\sum_{n=1}^M
s_{n}\left(  t\right)  ~g_{n}~\sigma_{z}.
\end{equation}
This is the case of pure dephasing noise.  The more general case
$H=-B_{0}\sigma_{z}/2-\sum_ns_n\left(  t\right)  \vec{g}\cdot\vec{\sigma}/2$ can also
be treated by the same method \cite{cheng}.  Using Eq. \ref{eq:tij}, we have
the $3\times3$ matrix
\begin{equation}
\label{eq:T_ij}
\begin{aligned}
T_{ij}\left(  \{s_n\},\Delta t\right)   &  =\frac{1}{2}\text{Tr~}\left[  \sigma
_{i}e^{i\left(  B_{0}+\sum s_{n}g_{n}\right)  \sigma_{z}\Delta t/2}\sigma
_{j}e^{-i\left(  B_{0}+\sum s_{n}g_{n}\right)  \sigma_{z}\Delta t/2}\right]\\
&  =\left\{\exp\left[  iL_{z}\left(  B_{0}+\sum_{n}s_{n}g_{n}\right)\Delta t \right]\right\}_{ij} ,
\end{aligned}
\end{equation}
where $L_{z}$ is the usual angular momentum matrix: $\left(  L_{z}\right)
_{xy}=-i$, $\left(  L_{z}\right)_{yx}=i$, and all other $\left(
L_{z}\right)  _{ij}=0.$ \ Eq. \ref{eq:T_ij} can be derived by direct calculation
or by noting that $\exp\left(  i\sigma_{z}\theta/2\right)  $ is a rotation by
$\theta$ about the $z$-axis in spin space.  Substituting Eqs. \ref{eq:w1} and
\ref{eq:T_ij} into Eqs. \ref{eq:gamma} and then using Eq. \ref{eq:Hq} we have
the quasi-Hamiltonian for this problem:
\begin{equation}
 H_{q}=\sum\limits_{n=1}^{M}\left(  -i\gamma_{n}+i\gamma_{n}\tau_{nx}
-g_{n}\tau_{nz}L_{z}\right)  -B_{0}L_{z}. \label{eq:qh}%
\end{equation}
Note $s_n$ in Eq. \ref{eq:T_ij} are replaced by $\tau_{nz}$ due to the first order expansion.

\subsection{Single Fluctuator}

We first consider the case of a single TLS, so that $M=1$ and $N_{c}=2$.
 This simple case illustrates all the essential mathematical features of the
method and the generalization to many independent TLS is almost immediate.
 We now have
\[
 H_{q}=-i\gamma+i\gamma\tau_{x}-g\tau_{z}L_{z}-B_{0}L_{z}.
\]
The problem of qubit evolution has been reduced to the diagonalization of the
$6\times6$ matrix $H_{q}$.  This is much simplified by the fact that
$ \left[  L_{z},H_{q}\right]  =0$ so the problem reduces to a set of 3
smaller problems for $L_{z}=0,\pm1$.  In these smaller problems no
manipulations more complicated than diagonalizing a $2\times2$ matrix are
required.  We treat the smaller blocks in turn.

1. $L_{z}=0$.  The $2\times2$ block of the quasi-Hamiltonian $H_{q}$ is
\[
 H_{q}\left(  L_{z}=0\right)  =\left(  -i\gamma+i\gamma\tau_{x}\right)
\]
There are $2$ eigenvalues and right eigenfunctions that satisfy
\[
H_{q}\left(  L_{z}=0\right)  \left\vert \Psi_{L_{z}=0}\right\rangle
=\omega_{L_{z}=0}\left\vert \Psi_{L_{z}=0}\right\rangle .
\]
We label them by $s=\pm1$.  The right eigenfunctions and eigenvalues are
\begin{align*}
\left\vert \Psi_{L_{z}=0}\left(  s\right)  \right\rangle  &  =\left(  \frac
{1}{\sqrt{2}}\right)
\begin{pmatrix}
0\\
0\\
1
\end{pmatrix}
~\otimes
\begin{pmatrix}
1\\
s
\end{pmatrix}
~\\
\omega_{L_{z}=0}\left(  s\right)   &  =-i\gamma+is\gamma
\end{align*}

2. $L_{z}=+1$.  The $2\times2$  block is
\begin{equation}
H_{q}\left(  L_{z}=1\right)  =-i\gamma+i\gamma\tau_{x}-g\tau_{z}-B_{0}
\label{eq:h1}
\end{equation}
The right eigenfunctions and eigenvalues for $s=\pm1$ are
\begin{align*}
\left\vert \Psi_{L_{z}=1}\left(  s\right)  \right\rangle  &  =C
\begin{pmatrix}
1/\sqrt{2}\\
i/\sqrt{2}\\
0
\end{pmatrix}
\otimes~
\begin{pmatrix}
i\gamma\\
g+s\sqrt{g^{2}-\gamma^{2}}
\end{pmatrix}
;~\\
\omega_{L_{z}=1}\left(  s \right)   &  =-i\gamma
-B_{0}+s\sqrt{g^{2}-\gamma^{2}}.
\end{align*}
Here $C^{-2}=2s\sqrt{g^2-\gamma^2}~(g+s\sqrt{g^{2}-\gamma^{2}})$.  The corresponding left eigenvectors $\left<\Psi_{L_z=1}\right|$ is simply the transpose of $\left|\Psi_{L_z=1}(s)\right>$ and $\left\langle \Psi_{L_{z}=1}\left(  s\right)
|\Psi_{L_{z}=1}\left(  s^{\prime}\right)  \right\rangle =\delta_{s,s^{\prime}
}$.  

3. \thinspace$L_{z}=-1$. 
\[
H_{q}\left(  L_{z}=1\right)  =-i\gamma+i\gamma\tau_{x}+g\tau_{z}+B_{0}
\]
Comparison to Eq. \ref{eq:h1} shows that the eigenvalues and eigenvectors for
$L_{z}=-1$ are obtained from the $L_{z}=1$ case by the substitutions $\left(
1/\sqrt{2},i/\sqrt{2},0\right)  \rightarrow\left(  1/\sqrt{2},-i/\sqrt{2},0\right)$,
$g_{n}\rightarrow-g_{n}$, and $B_{0}\rightarrow-B_{0}$. 

We now perform the average over initial states and sum over final states.  We
assume that $P_{s}\left(  t=0\right)$ has its steady state values
$P_{1}=P_{-1}=1/2$, or $\left|i_f\right>=[1,1]/2$.  However, the average and sum are more conveniently performed by
taking a partial inner product with the state $\left\vert\sqrt{p_{s}}\right\rangle $ of the classical system
\begin{equation}
\left\vert\sqrt{p_{s}}\right\rangle =\frac{1}{\sqrt{2}}
\begin{pmatrix}
1\\
1
\end{pmatrix}
,
\end{equation}
which projects onto the quantum subspace.  The final $3\times3$ evolution
matrix in this subspace is:
\[
\overline{T}\left(  t\right)  =\left\langle\sqrt{p_{s}}\right\vert e^{-iH_{q}
t}\left\vert\sqrt{p_{s}}\right\rangle .
\]
Note this equivalence between  $\left<\sqrt{p_s}\right|\cdot\left|\sqrt{p_s}\right>$ and $\left<x_f\right|\cdot\left|i_f\right>$ can only be established if the initial states are uniformly distributed.

Using the eigendecomposition of $H_{q}$ we have
\begin{equation}
\overline{T}(t)=\sum_{s,L_{z}}\left\langle \sqrt{p_s}\right\vert
\left.  \Psi_{L_{z}}\left(  s\right)  \right\rangle ~e^{-i\omega\left(
L_{z},s\right)  t}\left\langle \Psi_{L_{z}}\left(  s\right)  \right.
\left\vert \sqrt{p_s}\right\rangle ,
\end{equation}
so we need to do a sum of 6 terms for each member of the matrix. We now
compute each member of $\overline{T}_{ij}\left(  t\right)  $ in turn. 

For $\overline{T}_{zz}(t)$ only the $L_{z}=0$, $s=1$ term contributes and we find
\begin{align*}
\overline{T}_{zz}(t)  &  =\exp\left[  -i\omega_{L_{z}=0}\left( s= 1\right)
~t\right] \\
&  =1,
\end{align*}
This is simply the obvious statement that for pure dephasing noise, $n_{z}$
does not decay.  In terms of the standard relaxation times, this says that
$T_{1}=\infty$. 

$\overline{T}_{xz}(t)=\overline{T}_{zx}(t)=\overline{T}_{yz}(t)=\overline{T}_{zy}(t)=0$.

For $\overline{T}_{xx}(t)$ and $\overline{T}_{yy}(t)$ only the $L_{z}=\pm1$ blocks
contribute.  After a straightforward calculation one finds
\[
\overline{T}_{xx}(t)=\overline{T}_{yy}(t)=\exp\left(  -\gamma t\right)  \cos
B_{0}t\left[  \cos\left(  \sqrt{g^{2}-\gamma^{2}}t\right)  +\frac{\gamma
}{\sqrt{g^{2}-\gamma^{2}}}\sin\left(  \sqrt{g^{2}-\gamma^{2}}t\right)
\right],
\]
and finally
\[
\overline{T}_{xy}(t)=-\overline{T}_{yx}(t)=\exp\left(  -\gamma t\right)  \sin
B_{0}t~\left[  \cos\left(  \sqrt{g^{2}-\gamma^{2}}t\right)  +\frac{\gamma
}{\sqrt{g^{2}-\gamma^{2}}}\sin\left(  \sqrt{g^{2}-\gamma^{2}}t\right)
\right]  .
\]
So, for example, if  $\mathbf{n}\left(  t=0\right)=\left(  1,0,0\right)$,
then we have
\begin{equation}
\mathbf{n}\left(  t\right)  =\left(  \cos B_{0}t,\sin B_{0}t,0\right)
\exp\left(  -\gamma t\right)  \left[  \cos\left(  \sqrt{g^{2}-\gamma^{2}
}t\right)  +\frac{\gamma}{\sqrt{g^{2}-\gamma^{2}}}\sin\left(  \sqrt
{g^{2}-\gamma^{2}}t\right)  \right]  .
\end{equation}
The first factor is the uniform precession, and the rest of the expression
gives the decay and non-uniform precession due to the TLS.

In the analysis below, it will be convenient to deal with the relaxation
function $\Gamma\left(  t\right)$ defined by
\begin{equation}
\overline{T}_{xx}(t)=\cos B_{0}t~\exp\left[-\Gamma\left(t\right)\right],
\end{equation}
so
\begin{equation}
\Gamma\left(  t\right)  =\gamma t-\ln\left[  \cos\left(  \sqrt{g^{2}
-\gamma^{2}}t\right)  +\frac{\gamma}{\sqrt{g^{2}-\gamma^{2}}}\sin\left(
\sqrt{g^{2}-\gamma^{2}}t\right)  \right]  \label{eq:gammaweak}
\end{equation}

\subsubsection{Weak Coupling}

This is the case $\gamma>g$.  Note that weak coupling (small $g$) is the same
thing as fast switching (large $\gamma$).  The arguments of the trigonometric
functions are imaginary and $\Gamma(t)$ is better written in terms of
hyperbolic functions:
\[
\overline{T}_{xx}(t)=\exp\left(  -\gamma t\right) \cos B_0t \left[
\cosh\left(  \sqrt{\gamma^{2}-g^{2}}t\right)  +\frac{\gamma}{\sqrt{\gamma
^{2}-g^{2}}}\sinh\left(  \sqrt{\gamma^{2}-g^{2}}t\right)  \right]
\]
and the behavior at long times $\left(t\gg1/\gamma\right)$ is given by
\begin{equation}
\overline{T}_{xx}(t)\simeq\frac{1}{2}\left(  1+\frac{\gamma}{\sqrt{\gamma^{2}-g^{2}}
}\right)\exp\left[  \left(  -\gamma+\sqrt{\gamma^{2}-g^{2}}\right)
t\right]  .
\end{equation}
Thus the dephasing rate is
\begin{equation}
\frac{1}{T_{2}}=\gamma-\sqrt{\gamma^{2}-g^{2}}.
\end{equation}
For the extreme weak coupling case $\gamma\gg g$ we find
\begin{equation}
\frac{1}{T_{2}}=\frac{g^{2}}{2\gamma},
\end{equation}
which is the standard result from perturbation (Redfield) theory. 

In the short time limit $t\ll1/\gamma$ we have
\begin{align}
\Gamma\left(  t\right)=\frac{1}{2}g^{2}t^{2}-\frac{1}{6}\gamma g^2t^3+O(t^4).
\end{align}
This reuslt is interesting: it shows that the envelope function initially
decays quadratically even for a single fluctuator.  We shall see below that
this behavior is completely generic.

\subsubsection{Strong Coupling}

When $\gamma<g$, one has
\begin{equation}
\overline{T}_{xx}=\exp\left(  -\gamma t\right)  \cos B_{0}t~\left[
\cos\left(  \sqrt{g^{2}-\gamma^{2}}t\right)  +\frac{\gamma}{\sqrt{g^{2}%
-\gamma^{2}}}\sin\left(  \sqrt{g^{2}-\gamma^{2}}t\right)  \right]
\label{eq:strongt}
\end{equation}
and at short times $t\ll1/\sqrt{g^2-\gamma^2}$ we have
\begin{equation}
\overline{T}_{xx}=\cos B_{0}t~\left[1-\frac{1}{2} g^{2} t^{2}+\frac{1}{3}\gamma g^2t^3+O(t^4)\right],
\end{equation}
and 
\begin{align}
\Gamma\left(t\right)=\frac{1}{2}g^{2}t^{2}-\frac{1}{3}\gamma g^2t^3+O(t^4).
\end{align}
 Note that when the coupling constant $g$ is increased past $\gamma$ the relaxation
rate saturates at $\gamma$.  At the same point the oscillation frequency
bifurcates into the two frequencies $B_{0}\pm g/\sqrt{g^{2}-\gamma^{2}}$. 

We stress that Eq. \ref{eq:strongt} gives a result that is exact at strong
coupling. 

\subsection{Many Fluctuators}

The quasi-Hamiltonian is given by Eq. \ref{eq:qh}.  Again we have $\left[
L_{z},H_{q}\right]  =0$.  The quasi-Hamiltonian for each value of $L_{z}$ is
a sum of operators acting on the individual fluctuators, which is a sign of
the fact that they are statistically independent: they do not "interact" with
one another.  Thus the generalization from the single fluctuator case is
almost immediate. We have 
\begin{align*}
\overline{T}_{zz}(t) = 1,
\end{align*}
which is a sign of pure dephasing, and
\begin{align}
\overline{T}_{xx}(t)& =\overline{T}_{yy}(t)=\exp\left(  -\sum_{n=1}^{M}\gamma
_{n}t\right)  \cos B_{0}t~
{\displaystyle\prod\limits_{n=1}^{M}}
\left[  \cos\left(  \sqrt{g_{n}^{2}-\gamma_{n}^{2}}t\right)  +\frac{\gamma
_{n}}{\sqrt{g_{n}^{2}-\gamma_{n}^{2}}}\sin\left(  \sqrt{g_{n}^{2}-\gamma
_{n}^{2}}t\right)  \right],\\
\overline{T}_{xy}(t) &  =-\overline{T}_{yx}(t)=\exp\left(  -\sum_{n=1}^{M}
\gamma_{n}t\right)  \sin B_{0}t~ 
{\displaystyle\prod\limits_{n=1}^{M}}
\left[  \cos\left(  \sqrt{g_{n}^{2}-\gamma_{n}^{2}}t\right)  +\frac{\gamma
_{n}}{\sqrt{g_{n}^{2}-\gamma_{n}^{2}}}\sin\left(  \sqrt{g_{n}^{2}-\gamma
_{n}^{2}}t\right)  \right].
\end{align}

\subsubsection{Weak Coupling}
If $g_{n}<\gamma_{n}$ for all $n$, then
\[
\Gamma\left(  t\right)  =\sum_{n=1}^{M}\gamma_{n}t-\sum_{n=1}^{M}\ln\left[
\cosh\left(  \sqrt{\gamma_{n}^{2}-g_{n}^{2}}t\right)  +\frac{\gamma_{n}}
{\sqrt{\gamma_{n}^{2}-g_{n}^{2}}}\sinh\left(  \sqrt{\gamma_{n}^{2}-g_{n}^{2}
}t\right)  \right]  ,
\]
If $t\gg1/\min_{n}\left(  \gamma_{n}\right)$, then we have
\begin{align*}
\Gamma\left(  t\right)\simeq  \sum_{n=1}^{M}\left(  \gamma_{n}-\sqrt{\gamma_{n}^{2}-g_{n}^{2}
}\right)  t,
\end{align*}
so that the decay is exponential at long times.  For extreme weak coupling
$g_{n}\ll\gamma_{n}$ for all $n$ then the Redfield result holds:
\begin{equation}
\frac{1}{T_2}  =\frac{1}{2}\sum_{n=1}^{M}\frac{g_{n}^{2}}{\gamma_{n}}.
\end{equation}
At short times $t\ll1/\max_{n}\left(  \gamma_{n}\right)$
\begin{align}
\Gamma\left(  t\right)=\frac{1}{2}\sum_{n=1}^{M}g_{n}^{2}t^{2}
-\frac{1}{6}\sum_{n=1}^{M} \gamma_{n}g_{n}^{2}t^{3}.
\end{align}
We get deviations from the quadratic behavior at times of order
\begin{equation}
t\sim\frac{\sum_{n=1}^{M}g_{n}^{2}}{\sum_{n=1}^{M}\gamma_{n}g_{n}^{2}}.
\end{equation}
Thus the dephasing behavior is essentially exponential rather than quadratic in the weak coupling region.

\subsubsection{Strong Coupling}

If $g_{n}>\gamma_{n}$ for all $n$, then we find
\begin{equation}
\Gamma(t)=\sum_{n=1}^{M}\gamma_{n}t- \sum_{n=1}^{M}\ln
\left[  \cos\left(  \sqrt{g_{n}^{2}-\gamma_{n}^{2}}t\right)  +\frac{\gamma
_{n}}{\sqrt{g_{n}^{2}-\gamma_{n}^{2}}}\sin\left(  \sqrt{g_{n}^{2}-\gamma
_{n}^{2}}t\right)  \right]  . \label{eq:manystrong}
\end{equation}
This equation is exact and represents a new result for many strong-coupling fluctuators.

At short times $t\ll1/\max_{n}\sqrt{g_{n}^{2}-\gamma_{n}^{2}}$
\begin{equation}
\Gamma\left(  t\right)\simeq\frac{1}{2}\sum_{n=1}^{M} g_{n}^{2} t^{2}-
\frac{1}{3}\sum_{n=1}^{M} g_{n}^{2}\gamma_{n}t^{3}
\end{equation}
We get deviations from the initial quadratic behavior at times of order
\begin{equation}
t\sim t_{s}=\frac{\sum_{n=1}^{M} g_{n}^{2} }{\sum_{n=1}^{M}\gamma_{n} g_{n}^{2}}.
\end{equation}

When $t>t_{s}$ the behavior is more complicated.  We write
\begin{equation}
\Gamma\left(  t\right)  =\sum_{n=1}^{M}\gamma_{n}t-\ln
{\displaystyle\prod\limits_{n=1}^{M}}
\left[  \frac{1}{2}\left(  e^{i\lambda_{n}t}+e^{-i\lambda_{n}t}\right)
+\frac{\gamma_{n}}{2i\lambda_{n}}\left(  e^{i\lambda_{n}t}-e^{-i\lambda_{n}
t}\right)  \right]
\end{equation}
where $\lambda_n=\sqrt{g^2_n-\gamma^2_n}$ and we need to evaluate the expression
\begin{equation}
U\left(  t\right)  =\left(  \frac{1}{2}\right)  ^{M}
{\displaystyle\prod\limits_{n=1}^{M}}
\left\vert r_{n}\right\vert
{\displaystyle\prod\limits_{n=1}^{M}}
\left[  e^{i\left(  \lambda_{n}t+\theta_{n}\right)  }+e^{-i\left(  \lambda
_{n}t+\theta_{n}\right)  }\right]  ,
\end{equation}
with $r_{n}=1-i\gamma_{n}/\lambda_{n}=\left\vert r_{n}\right\vert \exp\left(
i\theta_{n}\right)  ;$ $\left\vert r_{n}\right\vert =\left(  1+\gamma_{n}
^{2}/\lambda_{n}^{2}\right)  ^{1/2}$ and $\theta_{n}=\tan^{-1}\left(
-\gamma_{n}/\lambda_{n}\right)$.

To this end we note that $\sum_{n=1}^{M}s_{n}\left(  \lambda_{n}t+\theta
_{n}\right)  $ is the result of a random walk $_{{}}$with a large number of
stpes $M$.  In the long-time limit $t\gg t_{l}=1/\min_{n}~g_{n}$ the central
limit theorem gives
\begin{align}
U\left(  t\right)   &  =
{\displaystyle\prod\limits_{n=1}^{M}}
\left(  1+\gamma_{n}^{2}/\lambda_{n}^{2}\right)  \exp\left(  -\frac{t^{2}}
{2}\sum_{n=1}^{M} g_{n}^{2} \right)  \\
&  =\exp\left\{  -\sum_{n=1}^{M}\left[  
 \frac{g_{n}^{2}t^{2}}{2}-\frac{g_{n}^{2}}{\lambda_{n}^{2}}\right]
\right\}  .
\end{align}
This expresssion of course neglects any Poincar\'{e} recurrences that can
occur whenever $M$ is finite.  In terms of $\Gamma$ we have
\begin{equation}
\Gamma\left(  t\right)  =t\sum_{n=1}^{M}\gamma_{n}+\frac{t^{2}}{2}\sum
_{n=1}^{M} g_{n}^{2} ,
\end{equation}
so that there is a regime of Gaussian decay at long times when the coupling is strong.

\subsection{Broad-spectrum Noise}

Finally we consider the case when the noise does not satisfy either of the
inequalities $g_{n}\gtrless\gamma_{n}$ for all $n$.  The fluctuators can
still be divided into $M_{f}$ fast (weak coupling, $g_{n}<\gamma_{n}$) 
fluctuators and $M_{s}$ slow (strong coupling, $g_{m}>\gamma_{m}$)
fluctuators.  As we have seen, this is not a qualitative distinction -
rather it corresponds to a change in the analytic behavior of the eigenvalues.
 This does not spoil the solvability of the model.  We have
\begin{multline}
\Gamma\left(  t\right)  =\sum_{n=1}^{M_{f}}\gamma_{n}t-\sum_{n=1}^{M_{f}
}\ln\left[  \cosh\left(  \sqrt{\gamma_{n}^{2}-g_{n}^{2}}t\right)
+\frac{\gamma_{n}}{\sqrt{\gamma_{n}^{2}-g_{n}^{2}}}\sinh\left(  \sqrt
{\gamma_{n}^{2}-g_{n}^{2}}t\right)  \right] \label{eq:bsnstart}\\
  +\sum_{m=1}^{M_{s}}\gamma_{m}t-
\sum\limits_{m=1}^{M_{s}}\ln
\left[  \cos\left(  \sqrt{g_{m}^{2}-\gamma_{m}^{2}}t\right)  +\frac{\gamma
_{m}}{\sqrt{g_{m}^{2}-\gamma_{m}^{2}}}\sin\left(  \sqrt{g_{m}^{2}-\gamma
_{m}^{2}}t\right)  \right]  ,
\end{multline}
and in the short time limit $t\ll t_{s}$:
\begin{equation}
\Gamma\left(  t\right)  \approx\frac{t^{2}}{2}\left[  \sum_{n=1}^{M_{f}
}g_{n}^{2} +\sum_{m=1}^{M_{s}} g_{m}^{2} \right]  ,
\end{equation}
while in the long time limit $t\gg t_{l}$:
\begin{equation}
\Gamma\left(  t\right)  \approx t\sum_{n=1}^{M_{f}}\left(  \gamma_{n}
-\sqrt{\gamma_{n}^{2}-g_{n}^{2}}\right)  +\frac{t^{2}}{2}\sum_{m=1}^{M_{s}
} g_{m}^{2}
\end{equation}
and the slow fluctuators will dominate at long times.  This result is
consistent with those obtained in Refs. \cite{paladino} and \cite{galperin} in
more specific models.

\section{Comparison to Approximate Solutions}

The most usual way to characterize noise is by its power spectrum.  Since our
noise sources satisfy Poisson statistics and are independent of each other, we
have
\begin{equation}
\overline{s_{m}\left(  t\right)  s_{n}\left(  t^{\prime}\right)  }=\delta
_{mn}\exp\left(  -2\gamma_{n}\left\vert t-t^{\prime}\right\vert \right),
\end{equation}
and the time auto-correlation function of the noise is
\begin{equation}
\sum_{mn}\overline{b_{mz}\left(  t\right)  b_{nz}\left(  t^{\prime}\right)
}~=\sum_{mn}g^{2}\overline{s_{m}\left(  t\right)  s_{n}\left(  t^{\prime
}\right)  }=g^{2}\sum_{n}\exp\left(  -2\gamma_{n}\left\vert t-t^{\prime
}\right\vert \right)  .
\end{equation}
The power spectrum is obtained by taking the Fourier transform:
\begin{equation}
S\left(  \omega\right)  =\frac{1}{2\pi}\int_{-\infty}^{\infty}dt~\overline
{b_{z}\left(  t\right)  b_{z}\left(  0\right)  }~e^{-i\omega t}.
\end{equation}
For our case this is
\begin{equation}
S_{cl}\left(  \omega\right)  =\frac{1}{\pi}\sum_{n=1}^{M}g_{n}^{2}
\frac{2\gamma_{n}}{4\gamma_{n}^{2}+\omega^{2}}:
\end{equation}
each individual fluctuator follows Poisson statistics and has a Lorentzian
power spectrum.  $S_{cl}\left(  \omega\right)  $ is an even function of
frequency;  this is probably the main limitation of our classical model, as
quantum noise is asymmetric in frequency at low temperatures \cite{girvin}.
 In the continuum limit, we find
\begin{equation}
S_{cl}\left(  \omega\right)  =\frac{1}{\pi}\int_{0}^{\infty}dg~\int
_{0}^{\infty}~d\gamma~p\left(  g,\gamma\right)  \frac{2g^{2}\gamma}
{4\gamma^{2}+\omega^{2}},
\end{equation}
where $p\left(  g,\gamma\right)  $ is the distribution of couplings and rates,
defined as
\begin{equation}
p\left(  g,\gamma\right)  =\sum_{n=1}^{M}\delta\left(  g-g_{n}\right)
~\delta\left(  \gamma-\gamma_{n}\right)  .
\end{equation}
When many fluctuators are superposed, we can obtain an arbitrary power
spectrum by the proper choice of $p\left(  g,\gamma\right)$.  Indeed, even
choosing $g_{n}=g_{0}$ independent of $n$ we have 
\begin{align}
S_{cl}\left(  \omega\right)   &  =\frac{1}{\pi}g_{0}^{2}\int_{0}^{\infty
}~d\gamma~p\left(  \gamma\right)  \frac{2\gamma}{4\gamma^{2}+\omega^{2}}\\
&  =\frac{1}{\pi}g_{0}^{2}~\int_{0}^{\infty}d\tau\cos\omega t\int_{0}^{\infty
}~d\gamma~p\left(  \gamma\right)  e^{-2\gamma t}.
\end{align}
Defining $p\left(  \gamma\right)  $ by $p\left(  g,\gamma\right)
=\delta\left(  g-g_{0}\right)  p\left(\gamma\right)$, this equation shows
that to obtain $p\left(  \gamma\right)  $ given $S_{cl}\left(\omega\right)
$, we first invert a Fourier cosine transform to obtain the original time
auto-correlation function and then $p\left(  \gamma\right)$ is proportional
to the the inverse Laplace transform of that.  We conclude that as long as
the only characterization of the noise is its power spectrum, then the results
given above provide an exact solution for any $S\left(  \omega\right)$.

We have already commented on the relation of the present solution to
perturbation (Redfield) theory.  The exact solution agrees with the
perturbative results when $g_{n}\ll\gamma_{n}$ for all $n$ and $t\gg1/\min_{n}\gamma_{n}$.

A more interesting approximation is the Gaussian approximation:
\begin{equation}
\Gamma_{G}\left(  t\right)  =\frac{t^{2}}{2}\int_{-\infty}^{\infty}
S_{cl}\left(  \omega\right)  \frac{\sin^{2}\omega t/2}{\left(  \omega
t/2\right)  ^{2}}d\omega.\label{eq:gammagauss}
\end{equation}
See, e.g., Ref. \cite{bergli} for a derivation.  This approximation is valid
when noise cumulants of third and higher order vanish.  For RTNs, this is not
the case - there are cumulants of all orders.  For a calculation of some of
these cunmulants, see Ref. \cite{lutchyn}.  Cumulants of order $n$ for a
single noise source are proportional to $g^{n}$, so we expect that the
Gaussian approximation will break down for large $g$. Qualitatively, the
behavior of $\Gamma_{G}\left(  t\right)  $ may be obtained by observing that
the function $\sin^{2}\left(  \omega t/2\right)  /\left(  \omega t/2\right)
^{2}$ acts largely as a filter function that passes frequencies $\omega<1/t,$
so
\begin{equation}
\Gamma_{G}\left(  t\right)  \approx t^{2}\int_{0}^{1/t}S_{cl}\left(
\omega\right)  d\omega.
\end{equation}
Furthermore, the total noise power is proportional to $\int_{-\infty}^{\infty
}S_{cl}\left(  \omega\right)  d\omega$.  For this integral to converge
(pathological cases apart) there must be an upper ($\omega_{uv})$ cutoff
frequency for $S_{cl}\left(  \omega\right)  $ and a lower ($\omega_{ir})$
frequency at which $S_{cl}\left(  \omega\right)  $ rolls over and becomes a
constant $S_{cl}\left(  0\right)$. Hence the asymptotic behaviors of
$\Gamma_{G}\left(  t\right)  $ are
\begin{equation}
\Gamma_{G}\left(  t\right)  \approx\left\{
\begin{array}
[c]{c}
t^{2}~\int_{0}^{\omega_{uv}}S_{cl}\left(  \omega\right)  d\omega
,~t\ll1/\omega_{uv}\\
t~S_{cl}\left(  0\right)  ,~t\gg1/\omega_{ir}.
\end{array}
\right.  \label{eq:asymptotic}
\end{equation}
There is an initial quadratic decrease of the signal and pure exponential
behavior at very long times.

It is now of interest to compare $\Gamma_{G}\left(  t\right)  $ with the exact
$\Gamma\left(  t\right)  $ for some interesting distributions $p\left(
g,\gamma\right)$.

\begin{figure} [tbp] 
\begin{center}
\includegraphics*[width=0.8\linewidth] {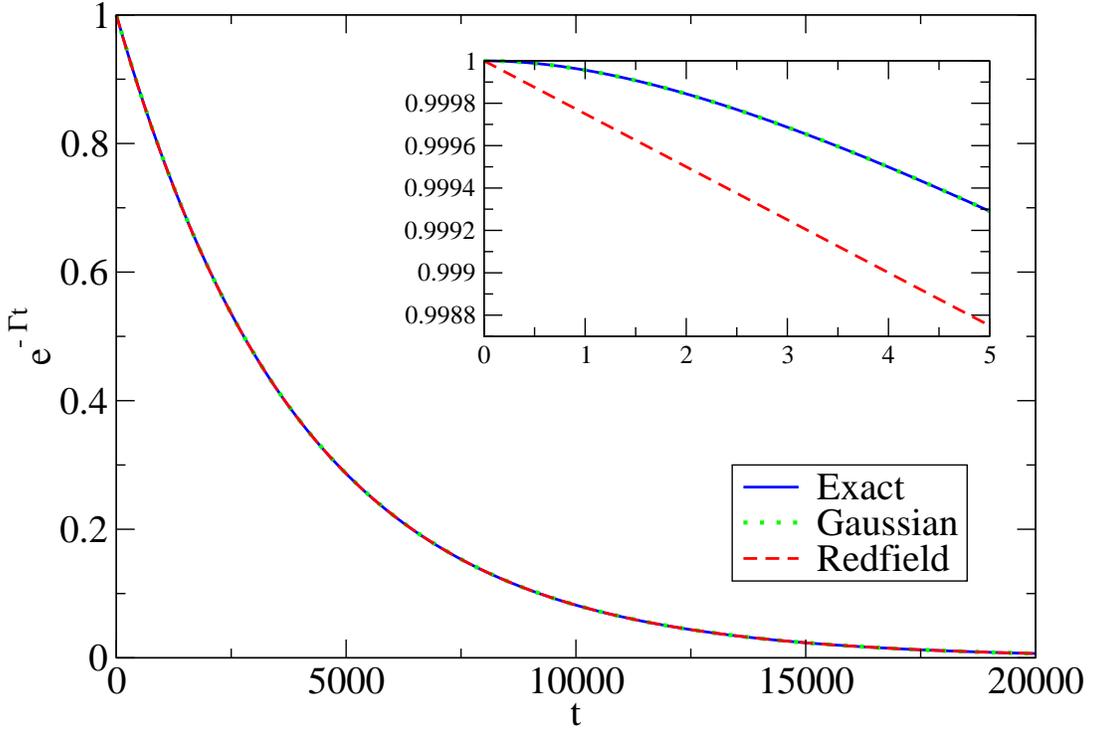}
\caption{(Color online) Envelope of the free induction decay signal $\exp\left[ -\Gamma\left(  t\right)  \right]  $  for weak coupling: g=0.01,$\gamma=0.2$.  The exact result is given by the blue solid line, Gaussian approximation by the green dashed line and the Redfield approximation by the red dashed line. The inset gives the short-time behavior.}
\label{fig:weak}
\end{center}
\end{figure}

\begin{figure} [tbp]
\begin{center}
\includegraphics*[width=0.8\linewidth] {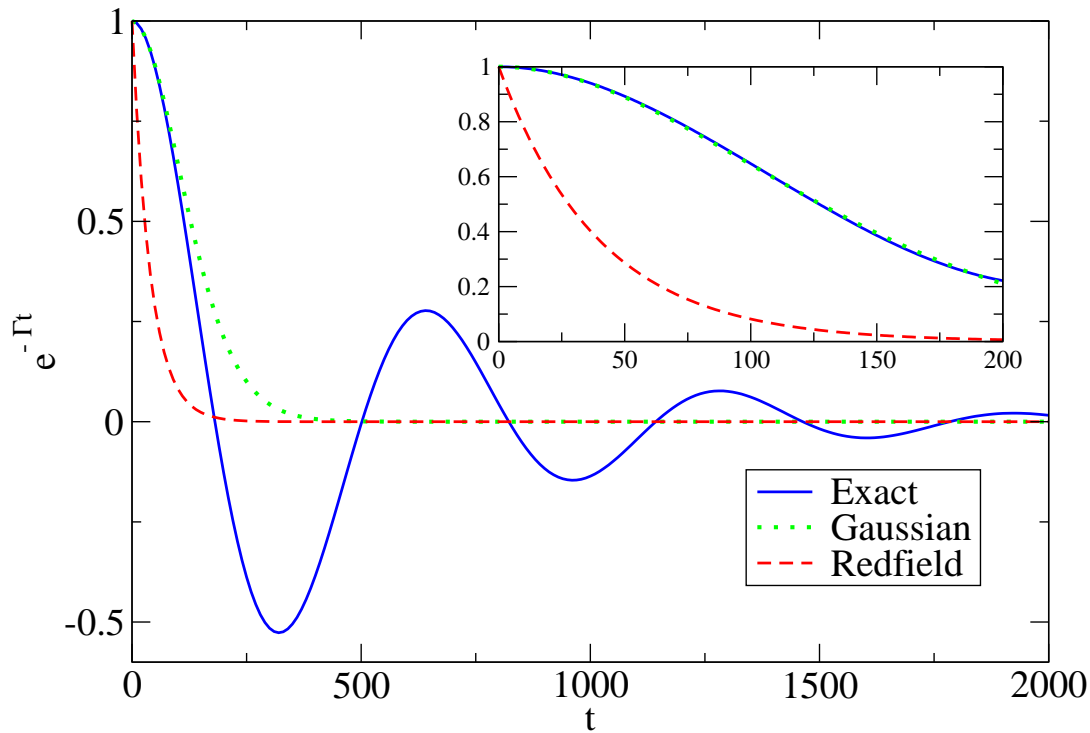}
\caption{(Color online) Envelope of the free induction decay signal $\exp\left[ -\Gamma\left(  t\right)  \right]  $  for strong coupling: g=0.01,$\gamma =0.002$.  The exact result is given by the blue solid line, Gaussian approximation by the green dashed line and the Redfield approximation by the red dashed line.  The inset gives the short-time behavior.}\label{fig:str}
\end{center}
\end{figure}

\begin {figure}[htbp]
    \includegraphics*[width=0.8\linewidth]{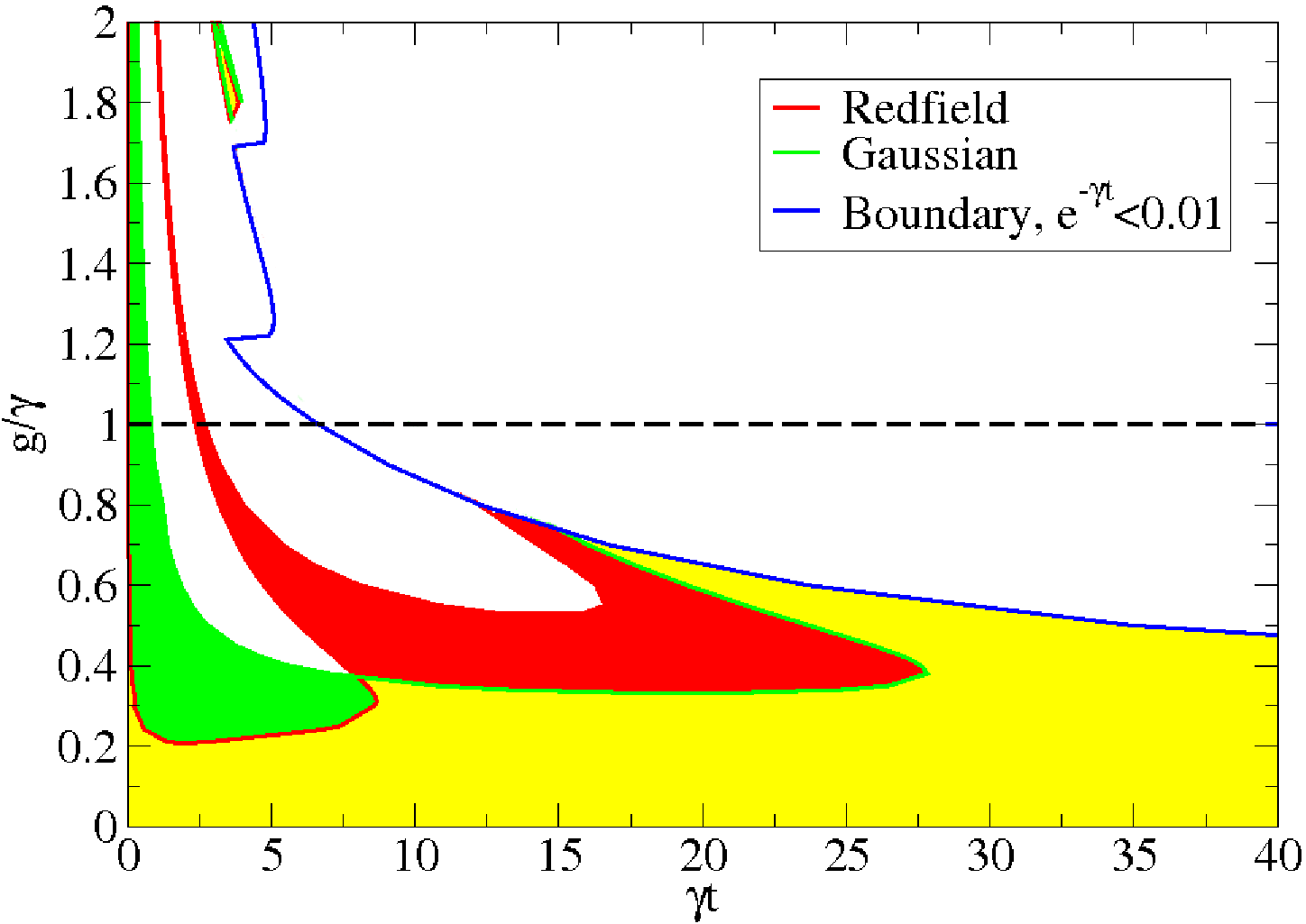}
    \caption{(Color online) Region of validity for Refield and Gaussian approximations. Envelope function $\exp[-\Gamma(t)]$ calculated from Gaussian and Redfield approximates are compared with the exact envelope. For this figure, we set $\gamma=1$. In the green and yellow region, the error of Gaussian approximation is smaller than $0.01$. In the red and yellow region, the error of Redfield approximation is smaller than $0.01$. The blue solid line is where the exact envelope dies off (smaller than $0.01$). The horizontal black dashed line separates the strong (top) and weak (bottom) coupling region. The red spike in the middle is due to the fact that the envelope calculated from Gaussian approximation crosses the exact envelope. The yellow island on top is due to the fact that both approximations die off much faster than the exact solution and the exact solution crosses zero multiple times, as seen in Fig.\ref{fig:str}. Note in Fig.\ref{fig:str}, $g/\gamma=5$, there will be $4$ islands for Gaussian approximation and $5$ islands for Redfield approximation. The blue boundary has zigzag shape in the strong coupling region. This is due to the oscillations of the exact envelope in the strong coupling region. }\label{fig:phase}
\end{figure}

For a single fluctuator, we have 
\begin{equation}
S_{cl}\left(  \omega\right)  =\frac{2}{\pi}g^{2}\frac{\gamma}{4\gamma
^{2}+\omega^{2}}
\end{equation}
so that the Gaussian approximation is \cite{galperin}
\begin{align}
\Gamma_{G}\left(  t\right)   &  =\frac{2g^{2}t^{2}}{2\pi}\int_{-\infty
}^{\infty}\frac{\gamma}{4\gamma^{2}+\omega^{2}}\frac{\sin^{2}\omega
t/2}{\left(  \omega t/2\right)  ^{2}}d\omega\\
&  =\frac{g^{2}}{4\gamma^{4}}\left(  e^{-2\gamma t}+2\gamma t-1\right)  .
\end{align}
Our results may now be compared with Redfield theory and the Gaussian
approximation (Eqs. \ref{eq:gammaweak}).  We give the decay function
$\exp\left[  -\Gamma\left(  t\right)  \right]  $ in Fig. \ref{fig:weak} and \ref{fig:str}.  
For weak coupling $g<\gamma$ [Fig. \ref{fig:weak}] Redfield theory works except at short times $t\ll1/\gamma$, while the Gaussian approximation is excellent at all times.  
For strong coupling $g>\gamma$ the situation is more complicated [Fig. \ref{fig:str}].  The
exact solution develops oscillations that are not present in the approximate
solution.  Again, Redfield theory is poor at short times, while the Gaussian
approximation is very good at these times, as already noted by other authors
\cite{bergli}.  At longer times both approximate solutions have little
resemblance to the exact solution.  We summarize the situation in Fig. \ref{fig:phase}.
 The areas of agreement (to within 1\%) are given by the white regions.
 Notice that the normalization is relative to the initial value of the
signal. 
At long times, the ratio of the exact results and the Gaussian
approximation can be much different from unity; however, the absolute value of
the signal is small and may be difficult to observe.  It is interesting to
note that the discrepancy between the approximate theory and exact theory is
oscillatory and is not well characterized as a "plateau".  This phenomenon is
more closely analyzed in Ref. \cite{zhou}. 

\section{Conclusions}

We have given a general formalism for the dissipative dynamics of an arbitrary
quantum system in the presence of a classical stochastic process.  It is
applicable to a very wide range of physical systems.  This method has several
virtues. It is linear, and the close analogy to Hamiltonian systems opens up
a toolbox of well-developed methods such as perturbation theory and mean-field
theory.  We applied the method to the problem of a single qubit in the
presence of TLS that give rise to pure dephasing 1/f noise and solved this
problem exactly.  This has been done before by the method of stochastic
differential equations \cite{galperin}.  However, that method depends on a
non-linear parameterization of the density matrix that is difficult to
generalize.  We anticipate that the method can be applied to other quantum
systems, such as an array of qubits, and also other kinds of noise. 

\begin{acknowledgments}
We would like to acknowledge useful discussions with S. N. Coppersmith, B.
Cheng, and D.T. Nghiem.  Financial support was provided by the National
Science Foundation, Grant Nos. NSF-ECS-0524253 and NSF-FRG-0805045, and by the
Defense Advanced Research Projects Agency QuEST program, and by th
Ministry of Science and Technology of China (Grant Nos. 2006CB921802 and
2006CB601002) and the 111 Project (Grant No. B07026).
\end{acknowledgments}

\appendix
\section{Properties of $T$ and $\overline{T}$}
\label{sec:app_A}

In this appendix we derive two properties of $T_{ij}$. 

1. $T$  is a real matrix.  This is shown as follows.
\begin{align*}
T_{ij}^{\ast} &  =\frac{1}{2}\text{Tr~}\sigma_{i}^{\ast}U^{\ast}\sigma
_{j}^{\ast}\left(  U^{\dag}\right)  ^{\ast}\\
&  =\frac{1}{2}\text{Tr }\sigma_{i}^{T}\left(  U^{-1}\right)  ^{T}\sigma
_{j}^{T}U^{T}\\
&  =\frac{1}{2}\text{Tr }U\sigma_{j}U^{-1}\text{ }\sigma_{i}\\
&  =\frac{1}{2}\text{Tr }\sigma_{i}U\sigma_{j}U^{\dag}\\
&  =T_{ij}%
\end{align*}

2. $T$ is an orthogonal matrix.  This is proved most simply by noting
that that the set of $2\times2$ Hermitian traceless matrices $A_{i}$ form a
3-dimensional real Hilbert space with inner product $\left(  A_{i}
,A_{j}\right)  =\frac{1}{2}$Tr$\left(  A_{i}A_{j}\right)  $.  The $\sigma
_{i}$ are a complete orthonormal basis for this space.  Here is the proof. 
\begin{align*}
T_{ij}T_{kj} &  =\frac{1}{4}\text{Tr~}\left(  \sigma_{i}U\sigma_{j}U^{\dag
}\right)  ~\text{Tr~}\left(  \sigma_{k}U\sigma_{j}U^{\dag}\right)  \\
&  =\frac{1}{2}\text{Tr~}\left(  U^{\dag}\sigma_{i}U\sigma_{j}\right)
\times\frac{1}{2}\text{Tr~}\left(  U^{\dag}\sigma_{k}U\sigma_{j}\right)  \\
&  =\left(  U^{\dag}\sigma_{i}U,\sigma_{j}\right)  \left(  U^{\dag}\sigma
_{k}U,\sigma_{j}\right)  \\
&  =\left(  U^{\dag}\sigma_{i}U,U^{\dag}\sigma_{k}U\right)  \\
&  =\frac{1}{2}\text{Tr~}\left(  U^{\dag}\sigma_{i}UU^{\dag}\sigma
_{k}U\right)  \\
&  =\frac{1}{2}\text{Tr~}\left(  \sigma_{i}\sigma_{k}\right)  \\
&  =\delta_{ik}.
\end{align*}
 The averaged quantity $\overline{T_{ij}\left(  t\right)  }$ is real since
the averaging is over real weights, but $\overline{T_{ij}\left(  t\right)  }$
is orthogonal only in trivial cases.

\end{document}